\documentclass[twocolumn,english,pra,showpacs,floatfix]{revtex4-1}
\usepackage[T1]{fontenc}
\usepackage[latin1]{inputenc}
\usepackage{graphicx}
\usepackage{amssymb}

\makeatletter

\usepackage{graphicx}

\usepackage{babel}
\makeatother

\newcommand \bea{\begin{eqnarray}}
\newcommand \eea{\end{eqnarray}}
\newcommand \ga{\raisebox{-.5ex}{$\stackrel{>}{\sim}$}}
\newcommand \la{\raisebox{-.5ex}{$\stackrel{<}{\sim}$}}

\begin{document}
\title{Inhomogeneous spin diffusion in traps with cold atoms}
\author{H. Heiselberg}
\email{heiselberg@mil.dk}
\affiliation{Joint Technology \& Innovation, DALO, Lautrupbjerg 1-5, DK-2750 Ballerup, Denmark}

\begin{abstract}
The spin diffusion and damped oscillations are studied in the collision of two spin polarized clouds of cold atoms with resonant interactions.
The strong density dependence of the diffusion coefficient leads to
inhomogeneous spin diffusion that changes from central to surface spin flow as the temperature
increases. The inhomogeneity and the smaller finite trap size
significantly reduce the spin diffusion rate at low temperatures. 
The resulting spin diffusion rates and spin drag at longer time scales
are compatible with measurements at low to high temperatures for resonant attractive interactions but are incompatible with a metastable ferromagnetic phase. 
This does not exclude that the colliding clouds can evolve into a repulsive initial state which subsequently decays during the bounce and the initial damped oscillations.
\end{abstract}
\maketitle

Measurements of spin diffusion in resonantly interacting trapped atomic gases \cite{Sommer} provide new understanding of the crossover physics and ferromagnetic (FM) phases. 
The transitions observed in radii and expansion energies versus repulsive interaction strength \cite{Jo} were interpreted as a FM Stoner transition \cite{Stoner} although disputed in Ref. \cite{Zhai}. 
A FM transition is predicted in several calculations \cite{Duine,Conduit,Pilati,Chang,FM} with
qualitatively similar transitions in observables but at a lower interaction strength as in the experiments.
Recent experiments on colliding spin
polarized gases and subsequent spin diffusion seem to exclude a metastable FM phase for resonant interactions \cite{Sommer}.
Since the spin diffusion in traps occurs on longer time scales than the initial bounce of the
spin separated clouds and subsequent damped oscillations, a FM phase 
initially as claimed by Taylor et al. \cite{Taylor} is not necessarily excluded. 

Resonant interactions are only characterized by an
infinite scattering length which can either be in the unitarity limit of strongly attractive
interactions (BCS-BEC crossover) or in the repulsive (FM phase). These
are also referred to as the lower and upper branches respectively of the multivalued scattering length (see e.g. \cite{FM}).
The upper branch is metastable and decays to the lower branch due to three-body interactions
\cite{Pekker}.

The diffusion coefficients measured in the traps seems to exclude a FM \cite{Sommer}
yet the diffusion rate exceeds the calculated values at central densities \cite{Bruun}
by up to an order of magnitude in particular at higher temperatures.
At temperatures well above the Fermi temperature $T\ga T_F$ the difference can be explained as a density inhomogeneity effect \cite{BP} arising because the spin diffusion is faster at the lower surface densities and as a result the spin diffusion circulates faster avoiding the nondiffusive core. The spin diffusion calculations will here be extended to low temperatures
$T\la T_F$ including the effects of density inhomogeneities. Comparison to the measured diffusion coefficients and the initial damped oscillations will then indicate whether a FM phase was present for resonant interactions. 

Defining the spin density $n_s=n_\uparrow-n_\downarrow$ and current ${\bf j}_s={\bf j}_\uparrow-{\bf j}_\downarrow$, the spin diffusion coefficient is defined from Fick's law. 
For inhomogeneous systems the diffusion equation leads to a modified form for Fick's law
\cite{BP} ${\bf j}_s=-D\chi \nabla (n_s/\chi)$, where $\chi=\partial n_s/\partial(\mu_\uparrow-\mu_\downarrow)$ is the spin susceptibility. Inserting the equation of continuity for spins $\partial_t n_s+\nabla {\bf j}_s =0$, we obtain the diffusion equation
for the spin diffusive mode
$n_s({\bf r},t)=e^{-t/\tau}n_s({\bf r})$
\begin{eqnarray} \label{DE}
\tau^{-1} n_s + \nabla (D\chi \nabla (n_s/\chi)) = 0 \,.
\end{eqnarray}
The axial spin diffusion rate $\tau^{-1}$ was calculated in Ref. \cite{BP} in the unitarity limit and at high temperatures, where 
$D\chi$ is independent of total density $n=n_\uparrow+n_\downarrow$. At low temperatures 
$D\chi\propto n^{5/3}$ in the unitarity limit whereas $D\chi\propto n$ 
in the dilute limit according to \cite{Bruun}; in both cases $\chi\propto n^{1/3}$. The various density dependences are conveniently parametrized by
$D\chi=D_0\chi_0 n^p/n_0^p$, where $p$ is a
general power dependence decreasing from $p=5/3$ at low temperatures to $p=0$ at high temperatures. $D_0$, $\chi_0$ and $n_0$ are the values in the trap center.
At low temperatures the atomic cloud density $n({\bf r})=n_0(1-r_\perp^2/R_\perp^2-z^2/R_z^2)^{3/2}$,
has a sharp surface at the Thomas-Fermi radii $R_i^2=2E_F/m\omega_i^2\sqrt{1+\beta}$. At the BCS-BEC crossover the universal 
constant is $\beta\simeq -0.54$ for unpolarized gases.

The diffusion equation (\ref{DE}) is precisely the one analyzed for collective modes in the case $p=1$, where analytical results exist \cite{mode}. 
The diffusive mode along the z-axis corresponds to the axial spin dipole mode with
$n_s/\chi\propto z$, which by insertion in the diffusion equation (\ref{DE}) gives
$\tau^{-1}=3D_0/R_z^2$ independent of trap deformation $\lambda=\omega_z/\omega_\perp$. 

The diffusion equation (\ref{DE}) results from using the variational principle on
\begin{eqnarray} \label{VE}
 \tau^{-1}_{var} = \frac{\int d^3r \, D\chi(\nabla(n_s/\chi))^2}{\int d^3r \, n_s^2/\chi } \,.
\end{eqnarray}
Varying the spin density $n_s$ on the right hand side gives an upper estimate for the spin diffusion rate, and its minimum determines both the spin density and diffusion rate.
It is experienced in transport calculations \cite{Pethick} that variational results are surprisingly accurate even
with the simplest ansatz that obeys conservation laws and symmetry. We shall also find this
in our case with $n_s/\chi\propto z$ (see Fig. (1)). We emphasize that the diffusion equation
(\ref{DE}) and the variational equation (\ref{VE}) differ in our low temperature case from Ref. \cite{BP} because $D\chi$ is not independent of density and also by the sharp density cutoff at the surface of the atomic cloud as will be explained shortly. Also the qualitative and quantitative change in diffusion is accurately captured in the variational calculations.

A variational calculation of Eq. (\ref{VE}) with the more general form
$n_s/\chi\propto z(1+c_zz^2/R_z^2+c_\perp r_\perp^2/R_\perp^2)$ results in a vanishing curvature $c_\perp=0$ in the radial direction for prolate traps, $\lambda\ll1$. Generally, we find that any such radial curvature increase the diffusion rate proportional to $c_\perp^2/\lambda^2$ due to the gradient in Eq. (\ref{VE}), and therefore it vanishes for finite prolate traps. The axial curvature $c_z$ is also independent of trap deformation when $\lambda\ll 1$, but does not vanish except at $p=1$, where it changes sign (see Fig. 1). 
A full variational calculation with $z^4$ and higher order contributions in $n_s/z\chi$ shows that these corrections are small and reduce the spin diffusion rate only slightly for $p\la 0$.
This is towards the high temperature limit where the cloud extends outside the Thomas-Fermi radii
anyway as discussed below.

The variational minimum yields the diffusion rate $\tau^{-1}_{var}\simeq (3D_0/R_z^2)I_p$,
which we write in terms on a dimensionless rate $I_p$ that contains the inhomogeneity of 
the trap diffusion. 
It generally depends on $p$ except as mentioned when $p=1$ where $I_p=1$ but is
independent of trap deformation when $\lambda\ll 1$. 
For prolate traps we find that
the ansatz $n_s\propto z\chi$ is a good approximation for positive $p$
as shown in Fig. 1, since the flow predominantly takes place along the symmetry axis near the center. 
Inserting this ansatz in Eq. (\ref{VE}), we get the (upper) estimate for the rate or in terms of the dimensionless "inhomogeneity" factor 
\bea \label{Ip}
I_p\equiv \frac{R_z^2}{3D_0\tau}
\simeq \frac{5\Gamma(4)\Gamma(3p/2+2)}{(3p+2)\Gamma(7/2)\Gamma(3p/2+5/2)} \,,
\eea
given in terms of the $\Gamma$-function. 
The importance of inhomogeneous spin diffusion is evident as shown in Fig. 1. The rate decreases with increasing power $p$ because the density and therefore the spin diffusion decrease toward the surface of the trap. 
Comparing temperatures $T\simeq T_F$ where $p\sim0$ to low temperatures where $p=5/3$, we find that the inhomogeneity factor is $\sim6$ times smaller at low temperatures. 

The diffusion rate at low temperatures is qualitatively different from the high temperature
case studied in Ref. \cite{BP}. At low temperatures the densities vanish outside the 
Thomas-Fermi radii
and therefore spin density and susceptibility also vanish. Consequently, the length scale in the radial direction is limited by $R_\perp$. This is contrary to high temperatures \cite{BP}, where the gas extends over
a length scale $l_i=\sqrt{2T/m\omega_i^2}$ but the spin diffusion extends over a much longer
radial length scale $\sim l_\perp/\sqrt{\lambda}$ because the diffusion coefficient increase with decreasing density. Therefore diffusion dominates in the surface and tail
of the distribution such that the spin flow circulates around the trap center.
As result the diffusion rate is increased by a factor $\lambda^{-2}$ as compared to the
low temperature result of Eq. (\ref{Ip}). However, because the diffusion eventually becomes
collisionless at very low densities in the tail, the rate is reduced to values similar as
the $p\sim0$ case above.

Using the spin diffusion coefficient from Ref. \cite{Bruun} for attractive interactions in the unitarity limit, we find good
agreement with experiment both at low and intermediate temperatures \cite{Sommer}. At high temperatures good agreement was also found $\tau^{-1}\simeq 10D_0/R_z^2$ when the rate was reduced due to collisionless surface region \cite{BP}. At low temperatures the diffusion
rate increases as $T^{-2}$ but is reduced by the smaller inhomogeneity factor as observed
in the experiments. 
The density dependence of the
diffusion coefficient, which implies a difference in inhomogeneity factor of $\sim6$ as seen in Fig. (1), is therefore important in order to describe both the magnitude and temperature dependence of the diffusion rate.

At the lowest temperatures accessible in the experiment of Sommer et al. \cite{Sommer}, superfluidity might start to appear in the center of the trap which is near spin equilibrium and therefore below the Clogston limit. As spin diffusion occurs in the center at low temperatures one may therefore look for increased spin diffusion and this way even determine the superfluid transition temperature and the Clogston limit.

\begin{figure}
\includegraphics[scale=0.6,angle=0]{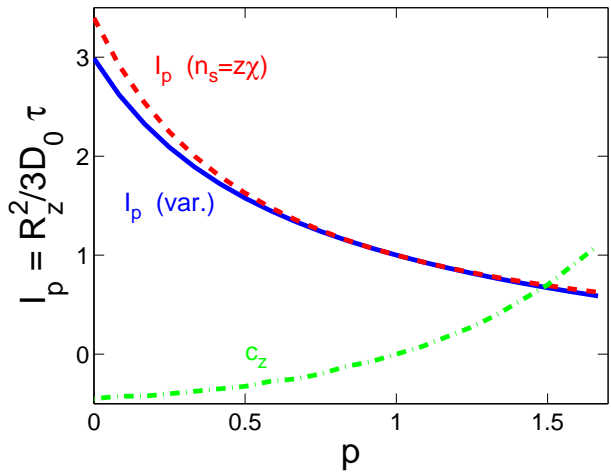}
\vspace{-0.5cm}
\caption{(Color online)
Spin diffusion rate versus the power of the density dependence for prolate traps, $\lambda\ll 1$ .
The inhomogeneity factor $I_p=R_z^2/3D_0\tau$
is calculated for $n_s/\chi\propto z$ (dashed curve) 
and variationally with $n_s/\chi=z(1+c_zz^2/R_z^2)$ (full curve), where $c_z$ (dash-dotted curve) has been varied to minimize the rate of Eq. (\ref{VE}).}
\end{figure}

 The spin drag coefficient $\Gamma_{sd}=\omega_z^2\tau$ measured in Ref. \cite{Sommer} 
can be expressed as
\begin{eqnarray}
 \frac{\hbar\Gamma_{sd}}{E_F} = \frac{2\hbar\sqrt{1+\beta}}{3I_pmD_0} \,.
\end{eqnarray}
It is independent of the trap frequencies as found in the experiments of Ref. \cite{Sommer}.
The inhomogeneity factor affects the spin drag and the diffusion rate by increasing the effective diffusion coefficient to $D=I_pD_0$. The minimum of the diffusion coefficient 
$D_0\simeq \hbar/m$ appears around medium temperature $T\sim 0.5T_F$ and determines the universal quantum limit to spin diffusivity in Fermi gases.

In the unitarity limit at low temperatures the isothermal conductivity $\kappa=2E_F/3n(1+F_0^s)$ measured in Ref. \cite{Sommer} 
is compatible with the lower branch where the spin-symmetric Landau parameter
$F_0^s=\beta\simeq -0.5$.
Likewise the spin susceptibility $\chi=2E_F/3n(1+F_0^a)$ is compatible with the lower branch
spin-asymmetric Landau parameter $F_0^a\simeq 1$ \cite{FM}. Both are incompatible with the upper branch
where $F_0^s$ is positive and $F_0^a$ negative.
F.ex. in the Jastrow-Slater approximation \cite{FM} $\beta=-0.54$ in the unitarity limit
on the lower branch but $\beta=2.93$ on the upper branch, and $F_0^a=-7\beta/5$.
Consequently the spin susceptibility is negative implying that the preferred ground state is FM with phase separated spins, which contradicts the diffusion and mixing observed on longer time scales.

In the experiments the two spin clouds are separated before ramping the
magnetic field on resonance. When the barrier is removed and the two clouds come into contact
the two-body wave function between unlike spins can evolve either into
the lower or upper branches corresponding to the
BCS-BEC crossover or ferromagnetic limit respectively. On the longer diffusive time scales 
the experimental setup either favors the
lower branch or the upper branch decays rapidly due to 3-body processes.
If the FM state relaxes from the upper
FM branch to the lower BCS branch before spin diffusion on the longer time scales, it would 
not necessarily contradict the claims in Ref. \cite{Taylor} that the two spin clouds are in a FM state at early times when they collide, bounce and undergo damped oscillations. 

In the recent work of Goulko et. al. \cite{Goulko}
the Boltzmann equation is solved at high temperatures off the unitarity limit. The bounce and oscillations of the two initially separated spin clouds and the initial dipole and breathing modes are calculated in detail. Viscous and diffusive relaxation are automatically included
but mean fields are ignored in the high temperature approximation. Therefore the bounce and damped oscillations depend on the scattering cross section only which is symmetric for positive and negative scattering lengths.
At lower temperatures we expect that the difference between the two scenarios of
attractive or repulsive interactions becomes important, i.e. whether the two spin clouds evolve into the lower or upper branch with attractive or repulsive mean fields respectively. 
The collisional cross section may be mainly responsible for the bounce and oscillation frequency but the sign should show up in the damping.
However, initially the two colliding spin
clouds are far from mechanical and chemical equilibrium 
and the subsequent damped oscillations are dominated by thermalization processes. The
damped oscillations seem not to discriminate between the evolution towards a repulsive or attractive initial state in the work by
Goulko et al. as they find good agreement without attractive or repulsive mean fields.
However, the calculations of Taylor et al. \cite{Taylor} favor repulsive mean fields.
Detailed analyses of the bounce, damped oscillations and diffusion are required at low temperatures before definite conclusions can be given. The colliding clouds are initially far from equilibrium and the thermalization affects the damped oscillations and diffusion.
Especially a more accurate determination of the size of the overlap zone and its temperature would be useful for a better understanding of the damped oscillations and diffusion, since their
rates are very temperature, density and spin density dependent.

In summary, the strong density and temperature dependence of the spin diffusion coefficient
leads to inhomogeneous spin diffusion that takes place in the center of the
cloud at low temperatures. This is opposite to the high temperature case where diffusion 
takes place in the surface and tail of the cloud circulating around the nondiffusive core.
The calculated inhomogeneity factor for resonant attractive interactions leads to good agreement with experiments at low and higher temperatures but
excludes a FM phase at the longer diffusive time scales. 
The initial bounce of the two colliding spin
clouds is far from mechanical and chemical equilibrium and the subsequent damped oscillations do not discriminate between the evolution towards a repulsive or attractive initial state.


\begin{thebibliography}{99}
\section*{References}
\bibitem{Sommer} A. Sommer, M. Ku, G. Roati, M. W. Zwierlein, Nature {\bf 472}, 201 (2011).
\bibitem{Jo} G-B. Jo et al., Science {\bf 325}, 1521-1524 (2009); see also comment ArXiv:0910.3419.
\bibitem{Stoner} E. Stoner, {\it Phil. Mag.} {\bf 15}, 1018 (1933).
\bibitem{Zhai} H. Zhai, Phys Rev. A {\bf 80}, 051605(R) (2009). 
\bibitem{Duine} R. A. Duine \& A. H. MacDonald, Phys. Rev. Lett. {\bf 95}, 230403 (2005).
\bibitem{Conduit} G. J. Conduit, B. D. Simons, Phys. Rev. Lett. {\bf 103}, 200403 (2009); arXIv:0907.3725.
\bibitem{Pilati} S. Pilati, G. Bertaina, S. Giorgini, M. Troyer, Phys. Rev. Lett. {\bf 105}, 030405 (2010).
\bibitem{Chang} S.-Y. Chang, M. Randeria, N. Trivedi, Proc. Natl. Acad. Sci. {\bf 108}, 51 (2011).
\bibitem{FM} H. Heiselberg, Phys. Rev. A {\bf 83},053635 (2011).
\bibitem{Taylor} E. Taylor, S. Zhang, W. Schneider, M. Randeria,
Phys. Rev. A {\bf 84}, 063622 (2011).
\bibitem{Pekker} D. Pekker et al., Phys. Rev. Lett. {\bf 106}, 050402 (2011).
\bibitem{Bruun} G. M. Bruun, New J. Phys. {\bf 13}, 035005 (2011);
G. M. Bruun, A. Recati, C. J. Pethick, H. Smith, S. Stringari, Phys. Rev. Lett. {\bf 100}, 240406 (2008).
\bibitem{BP} G. M. Bruun and C. J. Pethick, Phys. Rev. Lett. {\bf 107}, 255302 (2011).

\bibitem{mode} H. Heiselberg, Phys. Rev. Lett.  {\bf 93}, 040402 (2004). 

\bibitem{Pethick} G. Baym and C. J. Pethick, {\it Landau Fermi-Liquid Theory: Concepts and Applications}, Wiley \& Sons, 1991.

\bibitem{Goulko} O. Goulko, F. Chevy and C. Lobo, Phys. Rev. A {\bf 84}, 051605 (2011).


\end{thebibliography}
\end{document}